\documentclass{llncs}
\usepackage{amsmath, amsfonts, graphicx, booktabs, tabularx, multirow, siunitx, xcolor, soul, bm}
\usepackage{hyperref}

\graphicspath{{./figures/}}
\begin{document}

\title{3D Segmentation with Fully Trainable Gabor Kernels and Pearson's Correlation Coefficient\thanks{This paper was accepted by the International Workshop on Machine Learning in Medical Imaging (MLMI 2022). The final publication is available at Springer via \url{https://doi.org/10.1007/978-3-031-21014-3_6}}}
\author{Ken C. L. Wong, Mehdi Moradi}
 %index{Wong, Ken C. L.}
 %index{Moradi, Mehdi}
\institute{IBM Research -- Almaden Research Center, San Jose, CA, USA\\
\email{clwong@us.ibm.com}
}

\maketitle              % typeset the title of the contribution

\begin{abstract}
The convolutional layer and loss function are two fundamental components in deep learning. Because of the success of conventional deep learning kernels, the less versatile Gabor kernels become less popular despite the fact that they can provide abundant features at different frequencies, orientations, and scales with much fewer parameters. For existing loss functions for multi-class image segmentation, there is usually a tradeoff among accuracy, robustness to hyperparameters, and manual weight selections for combining different losses. Therefore, to gain the benefits of using Gabor kernels while keeping the advantage of automatic feature generation in deep learning, we propose a fully trainable Gabor-based convolutional layer where all Gabor parameters are trainable through backpropagation. Furthermore, we propose a loss function based on the Pearson's correlation coefficient, which is accurate, robust to learning rates, and does not require manual weight selections. Experiments on 43 3D brain magnetic resonance images with 19 anatomical structures show that, using the proposed loss function with a proper combination of conventional and Gabor-based kernels, we can train a network with only 1.6 million parameters to achieve an average Dice coefficient of 83\%. This size is 44 times smaller than the original V-Net which has 71 million parameters. This paper demonstrates the potentials of using learnable parametric kernels in deep learning for 3D segmentation.
\end{abstract}
\section{Introduction}

The convolutional layer and the loss function are two fundamental components in deep learning. Because of the success of conventional deep learning kernels, i.e., the kernels with weights directly trainable through backpropagation, advancements in deep learning architectures are mainly on combining existing layers and inventing new non-convolutional layers for better performance. On the other hand, traditional parametric kernels, such as the Gabor kernel, become less popular. In fact, the versatility of conventional deep learning kernels comes with the cost of enormous numbers of network parameters proportional to the kernel size. In contrast, parametric kernels are less versatile but more compact. In this paper, we focus on the Gabor kernel as it can provide features at different frequencies, orientations, and scales, which are important for image analysis.

Different frameworks have been proposed to benefit from Gabor kernels in deep learning for 2D image classification. In \cite{Conference:Sarwar:ISLPED2017}, some trainable conventional kernels were replaced by fixed Gabor kernels for more energy-efficient training. In \cite{Journal:Luan:TIP2018}, the trained conventional kernels in all convolutional layers were modulated by fixed Gabor filters to enhance the scale and orientation information. In \cite{Journal:Chen:TIS2019}, the first convolutional layer was composed of Gabor kernels with the sinusoidal frequency trainable by backpropagation. In \cite{Journal:Meng:Electronics2019}, the first convolutional layer was composed of Gabor kernels, where the sinusoidal frequencies and the standard deviations were trained by the multipopulation genetic algorithm. Although the results were promising, these frameworks require manual selections of some or all Gabor parameters. This diminishes the benefits of using deep learning and manual selections can be difficult for 3D problems.

For the loss functions for multi-class image segmentation, there are mainly three categories: pixel-based \cite{Conference:Ronneberger:MICCAI2015}, image-based \cite{Conference:Milletari:3DV2016,Conference:Salehi:MLMI2017,Conference:Berman:CVPR2018}, and their combinations \cite{Conference:Wong:MICCAI2018}. Pixel-based losses apply the same function on each pixel and their average value is computed. A popular pixel-based loss is the categorical cross-entropy \cite{Conference:Ronneberger:MICCAI2015}, which is robust to hyperparameters but may lead to suboptimal accuracy \cite{Conference:Berman:CVPR2018,Conference:Wong:MICCAI2018}. Image-based losses, such as the Dice loss \cite{Conference:Milletari:3DV2016}, Jaccard loss \cite{Conference:Berman:CVPR2018}, and Tversky loss \cite{Conference:Salehi:MLMI2017}, compute the losses from the prediction scores of all pixels in an image using statistical measures. The image-based losses can achieve better accuracy than the pixel-based ones, but are less robust to hyperparameters under certain situations \cite{Conference:Wong:MICCAI2018}. The pixel-based and image-based losses can complement each other by weighted combinations, though deciding the optimal weights is nontrivial. Therefore, it can be beneficial if we can find a loss function that is accurate, robust to hyperparameters, and does not require manual weight selections.

To address these issues, we propose two contributions in this paper. \textbf{I)} In 3D segmentation, it is difficult to manually decide the Gabor parameters, and this can introduce unnecessary kernels while the GPU memory is precious. To gain the benefits of Gabor kernels while keeping the advantages of automatic feature generation in deep learning, we propose a Gabor-based kernel whose parameters are fully trainable through backpropagation. By modifying the formulation of a 3D Gabor kernel, we improve the versatility of the proposed Gabor-based kernel while minimizing the memory footprint for 3D segmentation. To the best of our knowledge, this is the first work of using fully trainable Gabor kernels in deep learning for 3D segmentation. Moreover, this work shows the feasibility of using only parametric kernels for spatial convolution in deep learning. \textbf{II)} We propose a loss function based on the Pearson's correlation coefficient (PCC loss) which is robust to learning rate and provides high segmentation accuracy. Different from the Dice loss which is formulated by relaxing the integral requirement of the Dice coefficient (F$_1$ score), the Pearson's correlation coefficient is formulated for real numbers so no approximation is required. The PCC loss also makes the full use of prediction scores from both foreground and background pixels of each label, thus is more comprehensive than the categorial cross-entropy and Dice loss. Furthermore, in contrast to the Tversky and combinatorial losses, there are no additional weights to be manually decided. Experiments on 43 3D brain magnetic resonance images with 19 anatomical structures show that, with a proper combination of conventional and Gabor-based kernels, and the use of the PCC loss, we can train a network with only 1.6 million parameters to achieve an average Dice coefficient of 83\%. This is a 44 times reduction in size compared with the original V-Net which has 71 million parameters \cite{Conference:Milletari:3DV2016}.

\section{Methodology}

\subsection{Convolutional Layer with 3D Gabor-Based Kernels}

The real and imaginary parts of a 3D Gabor kernel can be represented as:
\begin{align}
\label{eq:Gabor_real_img}
G_{re} = A g(\mathbf{x}; \bm{\theta}, \bm{\sigma}) \cos(2 \pi f x' + \psi); \ \ G_{im} = A g(\mathbf{x}; \bm{\theta}, \bm{\sigma}) \sin(2 \pi f x' + \psi)
\end{align}
with $g(\mathbf{x}; \bm{\theta}, \bm{\sigma}) = \exp\left(-0.5\left((x'/\sigma_{x})^2 + (y'/\sigma_{y})^2 + (z'/\sigma_{z})^2\right)\right)$ the Gaussian envelope. $\bm{\theta} = (\theta_{x}, \theta_{y}, \theta_{z})$ are the rotation angles about the $x$-, $y$-, and $z$-axis, and $\bm{\sigma} = (\sigma_{x}, \sigma_{y}, \sigma_{z})$ are the standard deviations. $\mathbf{x} = (x, y, z)$ are the coordinates, and $(x', y', z')$ are the rotated coordinates produced by $\mathbf{R} = \mathbf{R}_{x}(\theta_{x})\mathbf{R}_{y}(\theta_{y})\mathbf{R}_{z}(\theta_{z})$, with $\mathbf{R}_{i}(\theta_{i})$ the basic rotation matrix about the $i$-axis. $A$ is the amplitude, $f$ is the frequency of the sinusoidal factor, and $\psi$ is the phase offset.

As we found that using a spherically symmetric Gaussian kernel provides similar results, we use $\sigma = \sigma_{x} = \sigma_{y} = \sigma_{z}$. Since $x'^{2} + y'^{2} + z'^{2} = \|\mathbf{x}\|^{2}$, now only $x'$ needs to be computed in (\ref{eq:Gabor_real_img}) and $\theta_{x}$ becomes unnecessary. Although existing works either use the real and imaginary part separately or only use the real part, to increase the versatility of the Gabor-based kernel while minimizing the memory footprint for deep learning in 3D segmentation, we add the two parts together and use different $A$ and $f$ for the real and imaginary parts:
\begin{align}
\label{eq:Gabor}
G_{DL} = g(\mathbf{x}; \sigma) \left(A_{re} \cos(2 \pi f_{re} x' + \psi) +  A_{im} \sin(2 \pi f_{im} x' + \psi)\right)
\end{align}
with $g(\mathbf{x}; \sigma) = \exp\left(-0.5\left((\|\mathbf{x}\|/\sigma)^2\right)\right)$. Compared with the conventional deep learning kernel with $k^3$ trainable parameters, with $k$ the kernel size, the proposed Gabor-based kernel only has eight parameters of $\{\sigma$, $\theta_{y}$, $\theta_{z}$, $A_{re}$, $A_{im}$, $f_{re}$, $f_{im}$, $\psi\}$. The conventional kernels can be replaced by $G_{DL}$ in a convolutional layer, and the necessary hyperparameters of that layer remain the same: $k$ and the number of output feature channels ($c_{out}$). Here $k$ is used to form the grid of coordinates $\mathbf{x}$. Therefore, the number of trainable parameters with $G_{DL}$ in a convolutional layer is $8 \times c_{in} \times c_{out}$, which is independent of $k$. As $G_{DL}$ is differentiable, the parameters can be updated through backpropagation.

\begin{figure}[t]
    \centering
    \begin{minipage}[t]{0.45\linewidth}
      \centering
      \includegraphics[width=1\linewidth]{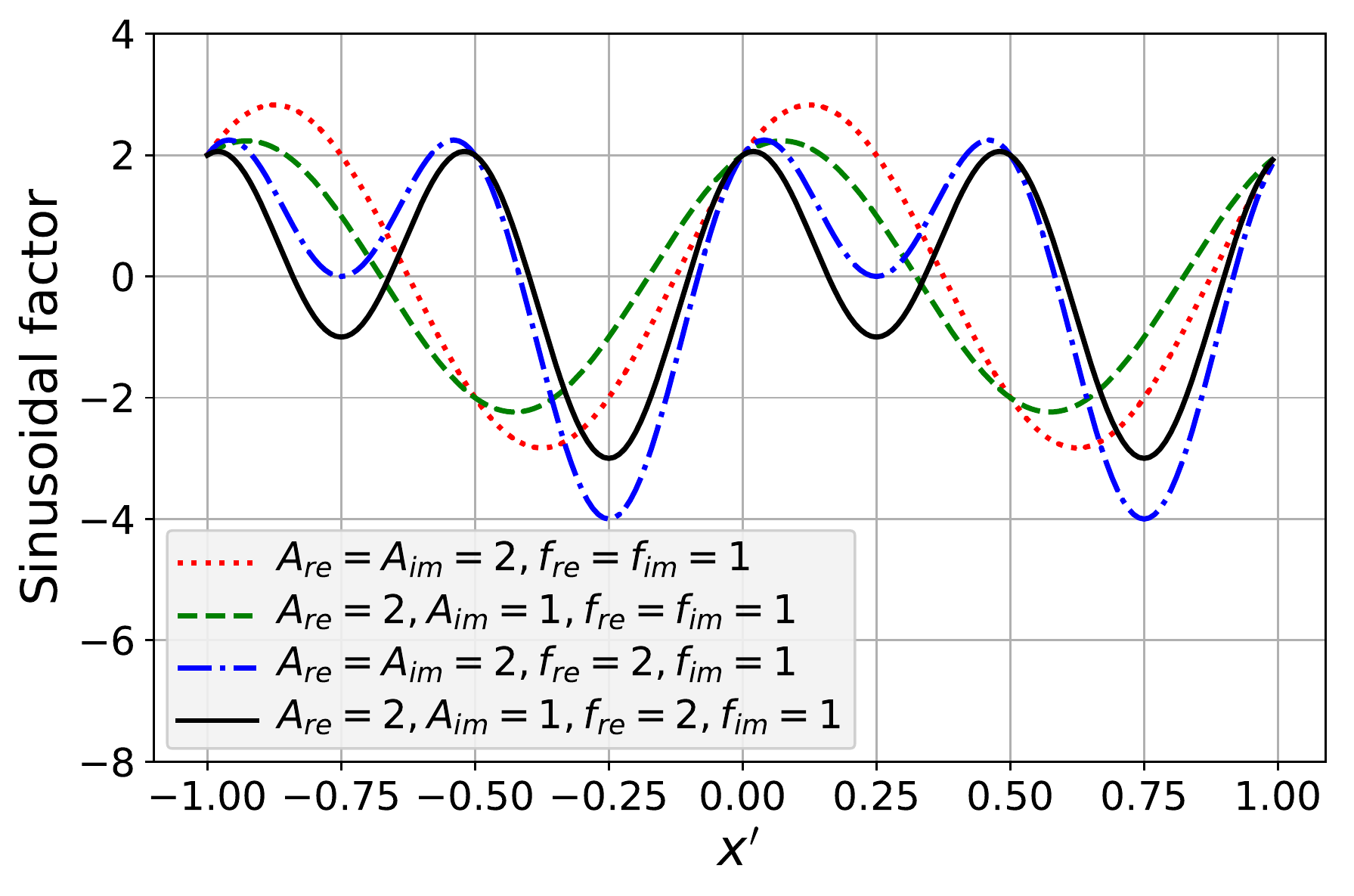}
      \centering{(a) Sinusoidal factors.}
    \end{minipage}
    \begin{minipage}[t]{0.45\linewidth}
      \centering
      \includegraphics[width=1\linewidth]{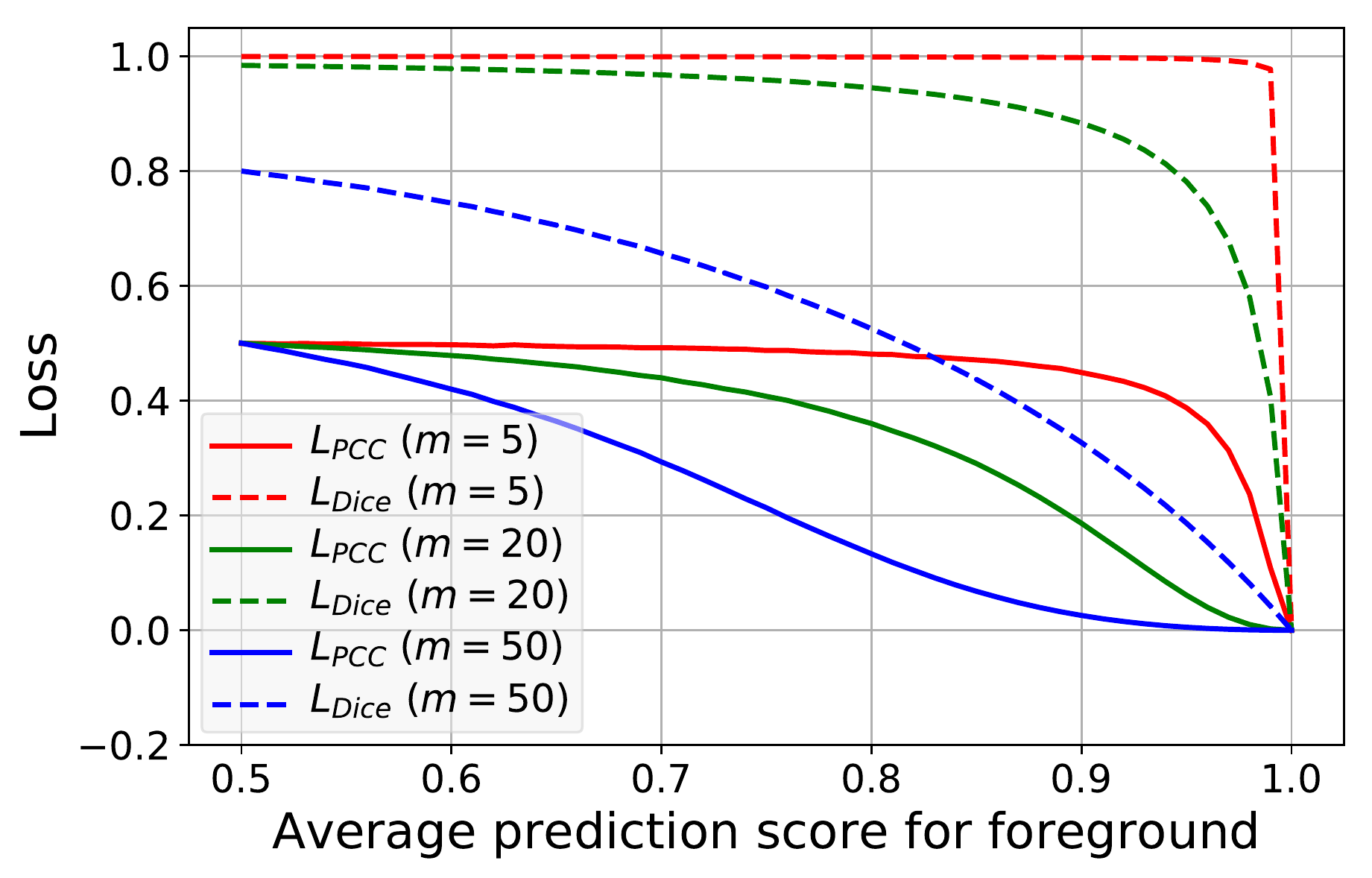}
      \centering{(b) Loss functions.}
    \end{minipage}
    \caption{(a) Sinusoidal factors with different amplitudes ($A$) and frequencies ($f$), $\psi=0$. (b) Comparison between $L_{PCC}$ and $L_{Dice}$ with different cubic object size $m$ ($m^3$ pixels).}
    \label{fig:demo}
\end{figure}

%Fig. \ref{fig:demo}(a) shows that (\ref{eq:Gabor}) can provide a variety of sinusoidal factors for better versatility.
Fig. \ref{fig:demo}(a) shows the characteristics of the sinusoidal factor in (\ref{eq:Gabor}). When $f_{re} = f_{im}$, the sinusoidal factor is similar to the common sinusoidal functions even with different $A_{re}$ and $A_{im}$. In contrast, more variations can be observed when $f_{re}$ and $f_{im}$ are different. Therefore, using different $A$ and $f$ for the real and imaginary parts can provide a larger variety of sinusoidal factors for better versatility.

% ------------------ figure -------------------------
% ---------------------------------------

\subsection{Loss Function with Pearson's Correlation Coefficient}

The Pearson's correlation coefficient ($r_{py} \in [-1, 1]$) measures the correlation between two variables. Given $N$ sample pairs $\{(p_1, y_1), \ldots, (p_N, y_N)\}$, we have:
\begin{align}
\label{eq:pcc}
r_{py} = \tfrac{\sum_{i=1}^{N}(p_i - \bar{p})(y_i - \bar{y})}{\sqrt{\left(\sum_{i=1}^{N}(p_i - \bar{p})^2\right)\left(\sum_{i=1}^{N}(y_i - \bar{y})^2\right)}}
\end{align}
where $\bar{p}$ and $\bar{y}$ are the sample means. Note that when $p_i$ and $y_i$ are binary, $r_{py}$ becomes the Matthews correlation coefficient which is known to be more informative than the F$_1$ score (Dice coefficient) on imbalanced datasets \cite{Journal:Chicco:BDM2017}.

For network training, we propose the PCC loss ($L_{PCC} \in [0, 1]$) as:
\begin{gather}
\label{eq:pcc_loss}
L_{PCC} = \mathbf{E}[1 - PCC_l]; \ \
PCC_l = 0.5 \left(\tfrac{\sum_{i=1}^{N}(p_{li} - \bar{p}_l)(y_{li} - \bar{y}_l)}{\sqrt{\left(\sum_{i=1}^{N}(p_{li} - \bar{p}_l)^2\right)\left(\sum_{i=1}^{N}(y_{li} - \bar{y}_l)^2\right)  + \epsilon}} + 1 \right)
\end{gather}
where $\mathbf{E}[\bullet]$ represents the mean value with respect to semantic labels $l$. $p_{li} \in [0, 1]$ are the network prediction scores, $y_{li} \in \{0, 1\}$ are the ground-truth annotations, and $N$ is the number of pixels of an image. $\epsilon$ is a small positive number (e.g., $10^{-7}$) to avoid the divide-by-zero situations, which happen when all $p_{li}$ or all $y_{li}$ are identical (e.g., missing labels). Therefore, $L_{PCC}$ = 0, 0.5, and 1 represent perfect prediction, random prediction, and total disagreement, respectively. As the means are subtracted from the samples in (\ref{eq:pcc}), both scores of the foreground and background pixels of each label contribute to $L_{PCC}$. Hence, a low $L_{PCC}$ is achievable only if both foreground and background are well classified. This is different from the Dice loss \cite{Conference:Milletari:3DV2016}:
\begin{gather}
\label{eq:dice_loss}
L_{Dice} = \mathbf{E}[1 - Dice_l]; \ \
Dice_l = \tfrac{2 \left(\sum_{i=1}^{N} p_{li} y_{li}\right) + \epsilon}{\left(\sum_{i=1}^{N} p_{li} + y_{li}\right)  + \epsilon}
\end{gather}
for which the background pixels do not contribute to the numerator.

Fig. \ref{fig:demo}(b) shows the comparison between $L_{PCC}$ and $L_{Dice}$ in a simulation study. Suppose that there is a cubic image of length 100 (i.e., 100$^3$ pixels) with a cubic foreground object of length $m$. To simulate a training process, $p_{li}$ in the foreground are drawn from a normal distribution whose mean and standard deviation change linearly from 0.5 to 1 and from 0.5 to 0, respectively, i.e., from $\mathcal{N}(0.5, 0.5^2)$ to $\mathcal{N}(1, 0)$. For those in the background, the distribution changes linearly from $\mathcal{N}(0.5, 0.5^2)$ to $\mathcal{N}(0, 0)$. The drawn $p_{li}$ are clipped between 0 and 1. Therefore, $p_{li}$ change from totally random to perfect scores. In Fig. \ref{fig:demo}(b), regardless of the object size, $L_{PCC}$ consistently starts at 0.5 and ends at 0, whereas the starting value of $L_{Dice}$ depends on the object size. Furthermore, when the object is small ($m$ = 5), the gradient of $L_{Dice}$ is very small and suddenly changes abruptly around the prediction score of 0.99. This means that a small learning rate is required when training with $L_{Dice}$ especially for small objects.

% ------------------ figure -------------------------
\begin{figure}[t]
    \centering
    \begin{minipage}[t]{0.95\linewidth}
      \centering
      \includegraphics[width=1\linewidth]{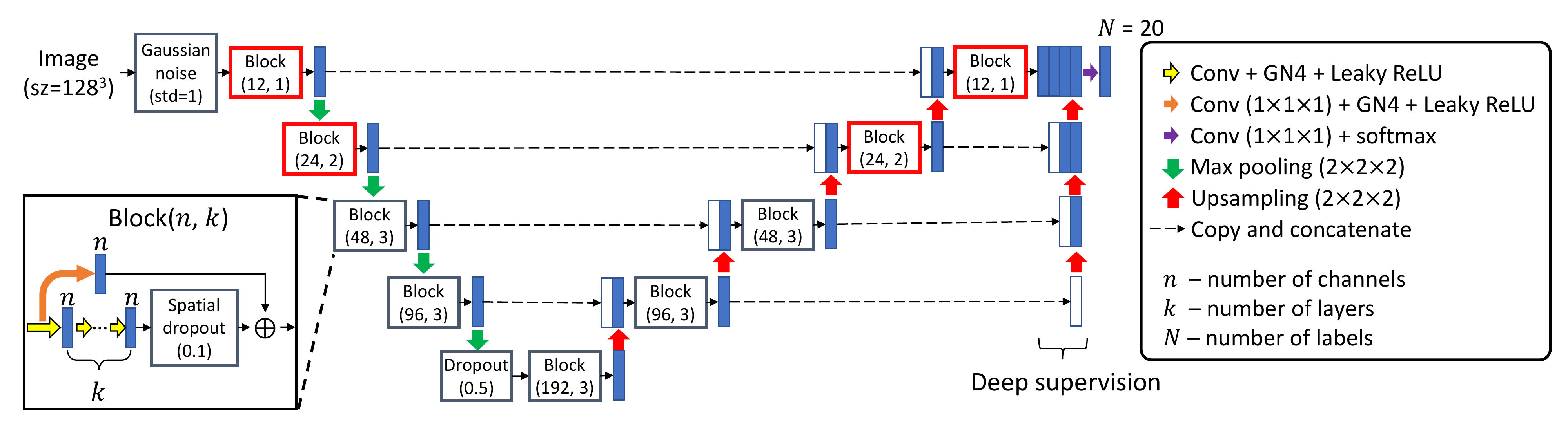}
    \end{minipage}
    \caption{Network architecture. Blue and white boxes indicate operation outputs and copied data. GN4 stands for group normalization with four groups of channels. The convolutional layers (Conv) of yellow arrows can comprise conventional kernels (3$\times$3$\times$3) or Gabor-based kernels (7$\times$7$\times$7). For the mixed-kernel models, the red blocks comprise the conventional kernels while the others comprise the Gabor-based kernels.}
    \label{fig:network}
\end{figure}
% ---------------------------------------

\subsection{Network Architecture}
\label{sec:network}

We modify the network architecture in \cite{Conference:Wong:MICCAI2018} which combines the advantages of low memory footprint from the V-Net and fast convergence from deep supervision (Fig. \ref{fig:network}). Each block in Fig. \ref{fig:network} comprises the spatial convolutional layers, which can be composed of conventional or Gabor-based kernels. Spatial dropout \cite{Conference:Tompson:CVPR2015} and residual connection \cite{Conference:He:ECCV2016} are used to reduce overfitting and enhance convergence. As the batch size is usually small for 3D segmentation because of memory requirements (e.g., one for each GPU), group normalization \cite{Conference:Wu:ECCV2018} is used instead of batch normalization for better accuracy, and four groups of channels per layer gave the best performance in our experiments.

For conventional kernels, the kernel size of 3$\times$3$\times$3 is used as it gave good results in the experiments. For Gabor-based kernels, although the numbers of trainable parameters are independent of the kernel size, a larger kernel size is more adaptive to different frequencies and scales, and a kernel size of 7$\times$7$\times$7 is chosen empirically. Different kernel combinations were tested (Section \ref{sec:experiments}), including models with only conventional kernels (conventional models), with only Gabor-based kernels (Gabor-based models), and with a mix of conventional and Gabor-based kernels (mixed-kernel models). As we want the mixed-kernel models to have small numbers of trainable parameters while achieving good performance, the conventional kernels are only used by the layers with fewer input and output channels, i.e., the red blocks in Fig. \ref{fig:network}.

\subsection{Training Strategy}

To avoid overfitting, image augmentation with rotation (axial, $\pm$\ang{30}), shifting ($\pm$20\%), and scaling ([0.8, 1.2]) was used, and each image had an 80\% chance to be transformed. The optimizer Nadam was used for fast convergence, and different learning rates and loss functions were tested in the experiments (Section \ref{sec:experiments}). Two NVIDIA Tesla V100 GPUs with 16 GB memory were used for multi-GPU training with a batch size of two and 300 epochs.

\section{Experiments}
\label{sec:experiments}

\subsection{Data and Experimental Setups}

A dataset of 43 3D T1-weighted MP-RAGE brain magnetic resonance images was used. The images were manually segmented by highly trained experts, and each had 19 semantic labels of brain structures. Each image was resampled to isotropic spacing, zero padded, and resized to 128$\times$128$\times$128.

Five dataset splits were generated by shuffling and splitting the dataset, each with 60\% for training, 10\% for validation, and 30\% for testing. The validation sets were used to choose the best epoch in each training. Three kernel combinations, including the conventional models, Gabor-based models, and mixed-kernel models (Section \ref{sec:network}), were tested with three loss functions of $L_{PCC}$, $L_{Dice}$, and categorical cross-entropy \cite{Conference:Wong:MICCAI2018}. Each of these nine combinations was tested with five learning rates (10$^{-4}$, 10$^{-3.5}$, 10$^{-3}$, 10$^{-2.5}$, 10$^{-2}$) on the five splits. Therefore, 225 models were trained. Because of the page limit, we only compare with the basic loss functions, and their combinations are not presented.

% ------------------ figure -------------------------
\begin{figure}[t]
    \centering
    \begin{minipage}[t]{0.32\linewidth}
      \scriptsize
      \centering
      \includegraphics[width=1\linewidth]{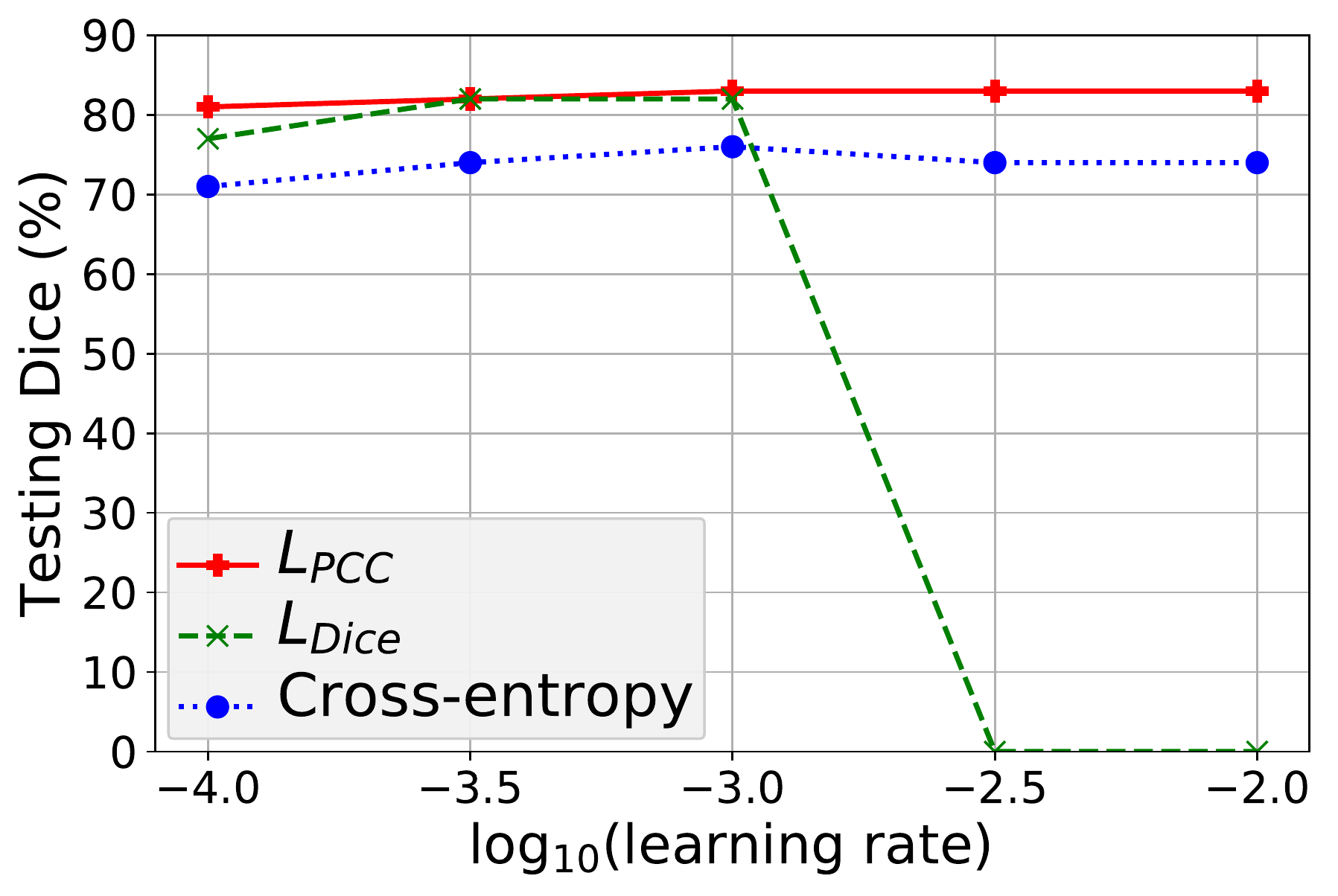}
      \centering{(a) Conventional.}
    \end{minipage}
    \begin{minipage}[t]{0.32\linewidth}
      \scriptsize
      \centering
      \includegraphics[width=1\linewidth]{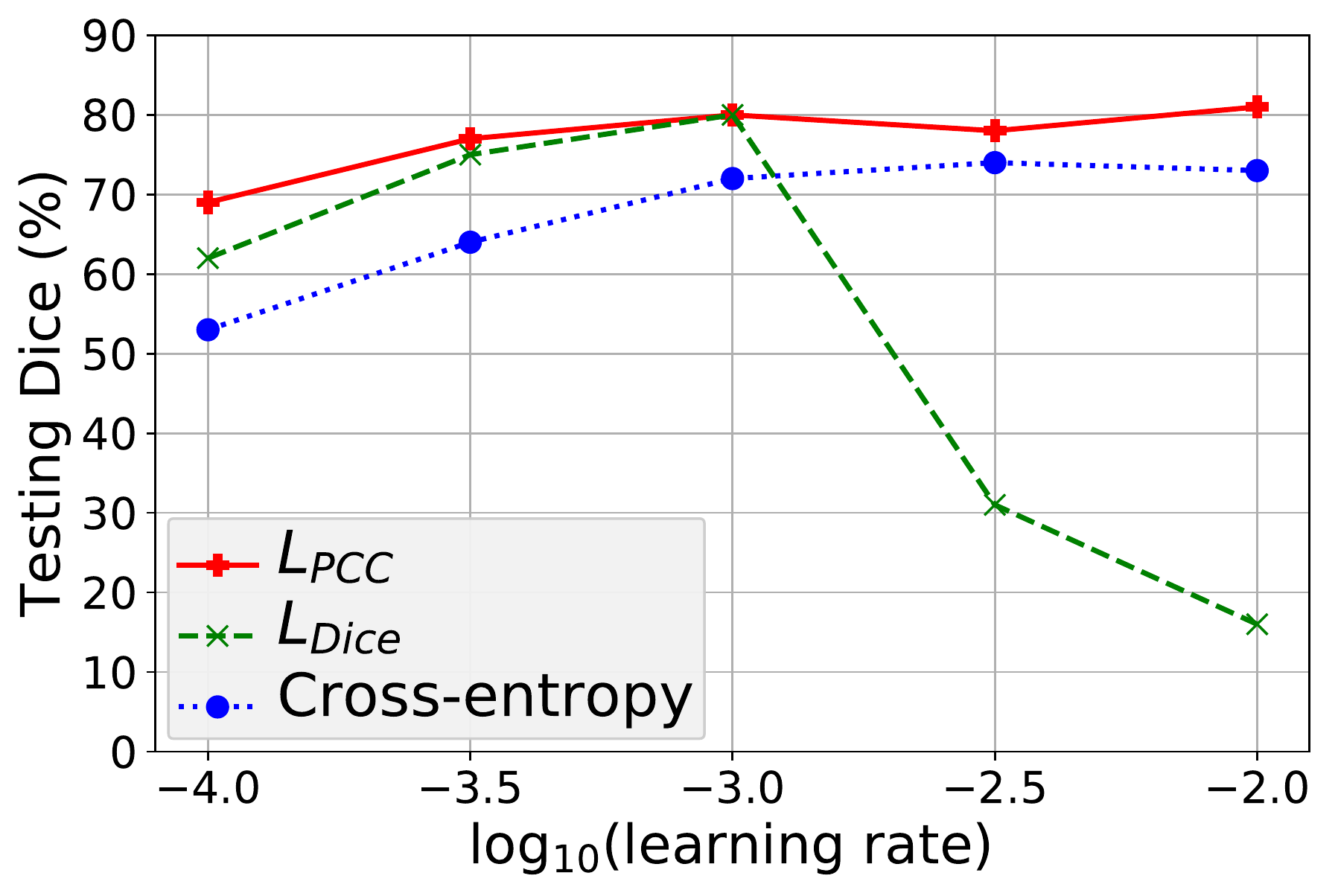}
      \centering{(b) Gabor-based.}
    \end{minipage}
    \begin{minipage}[t]{0.32\linewidth}
      \scriptsize
      \centering
      \includegraphics[width=1\linewidth]{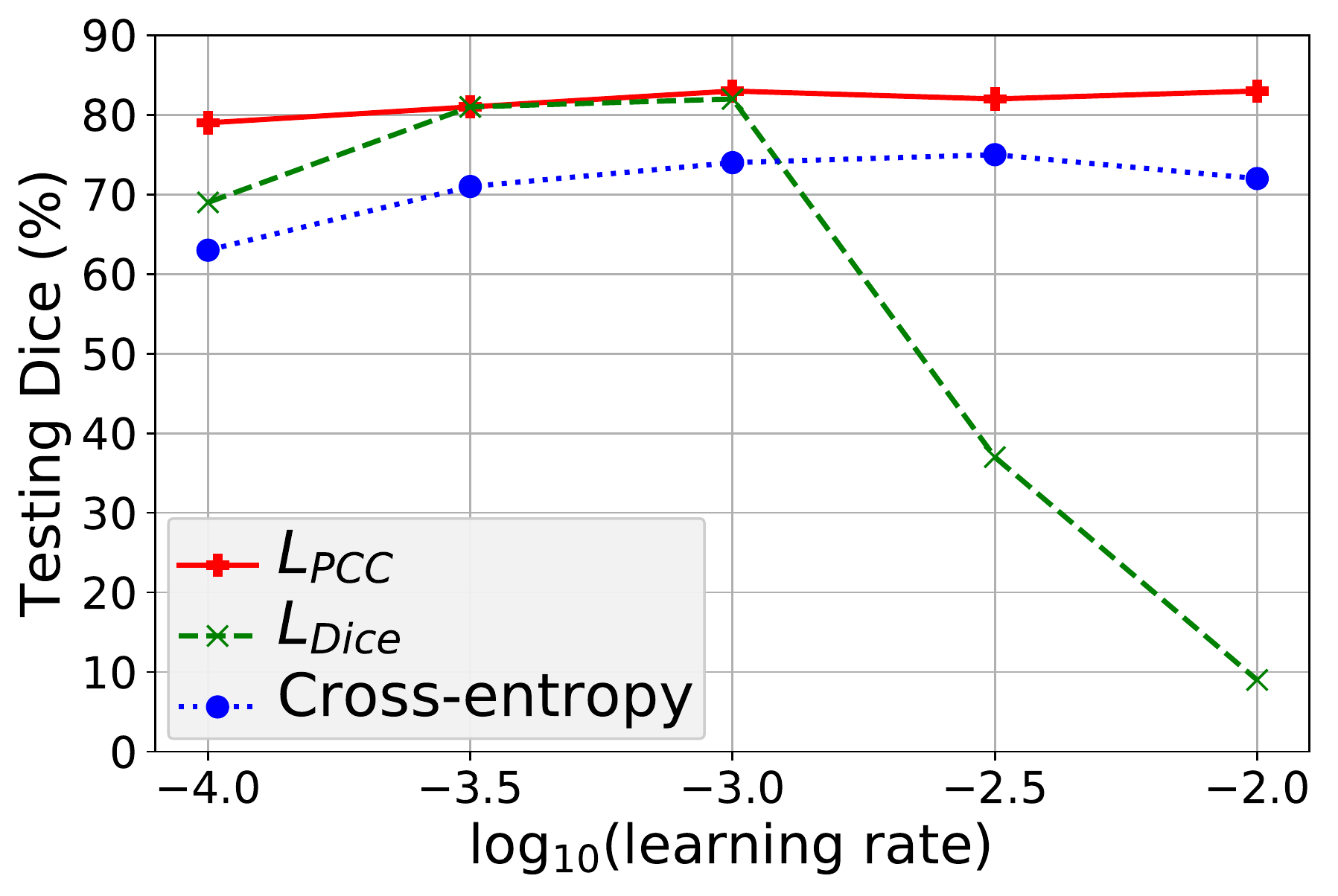}
      \centering{(c) Mixed-kernel.}
    \end{minipage}
    \caption{The robustness to learning rate of different loss functions with different kernel combinations. The value of each point is averaged from five experiments.}
    \label{fig:learning_rate}
\end{figure}
% ---------------------------------------

\subsection{Results and Discussion}

% ------------------ Table -------------------------
\begin{table*}[t]
\caption{Semantic brain segmentation at learning rate 10$^{-3}$. (a) Semantic labels and their relative sizes on average (\%). CVL represents cerebellar vermal lobules. (b) Testing Dice coefficients between prediction and ground truth averaged from five experiments (format: mean$\pm$std\%). The best results are highlighted in blue.}
\label{table:brain}

\smallskip
\fontsize{6}{7}\selectfont
\centering

\newcolumntype{R}{>{\raggedleft\arraybackslash}X}

\centering{(a) Semantic labels and their relative sizes on average (\%).}
\begin{tabularx}{\linewidth}{RllRllRllRll}
\toprule
1. & Cerebral grey & (50.24) & 2. & 3rd ventricle & (0.09) & 3. & 4th ventricle & (0.15) & 4. & Brainstem & (1.46) \\
\midrule
5. & CVL I-V & (0.39) & 6. & CVL VI-VII & (0.19) & 7. & CVL VIII-X & (0.26) & 8. & Accumbens & (0.07) \\
\midrule
9. & Amygdala & (0.21) & 10. & Caudate & (0.54) & 11. & Cerebellar grey & (8.19) & 12. & Cerebellar white & (2.06) \\
\midrule
13. & Cerebral white & (31.23) & 14. & Hippocampus & (0.58) & 15. & Inf. lateral vent. & (0.09) & 16. & Lateral ventricle & (2.11) \\
\midrule
17. & Pallidum & (0.25) & 18. & Putamen & (0.73) & 19. & Thalamus & (1.19) \\
\bottomrule
\medskip
\end{tabularx}

\centering{(b) Average testing Dice coefficients (mean$\pm$std\%) with respective to the ground truth.}
\begin{tabularx}{\linewidth}{lRlRlRlRlRlRlRl}
\toprule
\multicolumn{15}{c}{Conventional (4.99 million parameters)} \\
\midrule
\multirow{3}{0.18\linewidth}{$L_{PCC}$}
& 1. & {\color{blue}\textbf{88$\pm$1}} & 2. & {\color{blue}\textbf{80$\pm$2}} & 3. & {\color{blue}\textbf{85$\pm$1}} & 4. & {\color{blue}\textbf{91$\pm$0}} & 5. & {\color{blue}\textbf{84$\pm$0}} & 6. & 75$\pm$1 & 7. & 80$\pm$1 \\
& 8. & {\color{blue}\textbf{71$\pm$2}} & 9. & {\color{blue}\textbf{77$\pm$1}} & 10. & {\color{blue}\textbf{86$\pm$1}} & 11. & {\color{blue}\textbf{90$\pm$0}} & 12. & {\color{blue}\textbf{87$\pm$1}} & 13. & {\color{blue}\textbf{90$\pm$1}} & 14. & {\color{blue}\textbf{82$\pm$1}} \\
& 15. & {\color{blue}\textbf{65$\pm$1}} & 16. & {\color{blue}\textbf{91$\pm$0}} & 17. & {\color{blue}\textbf{81$\pm$1}} & 18. & {\color{blue}\textbf{88$\pm$1}} & 19. & {\color{blue}\textbf{90$\pm$0}}
& \multicolumn{4}{l}{\ {\color{blue}\textbf{Average: 83$\pm$0}}} \\
\midrule
\multirow{3}{0.18\linewidth}{$L_{Dice}$}
& 1. & 87$\pm$1 & 2. & {\color{blue}\textbf{80$\pm$2}} & 3. & {\color{blue}\textbf{85$\pm$0}} & 4. & {\color{blue}\textbf{91$\pm$1}} & 5. & 83$\pm$1 & 6. & 74$\pm$1 & 7. & 80$\pm$1 \\
& 8. & 70$\pm$1 & 9. & 76$\pm$2 & 10. & 85$\pm$1 & 11. & 89$\pm$0 & 12. & {\color{blue}\textbf{87$\pm$1}} & 13. & 88$\pm$1 & 14. & 81$\pm$1 \\
& 15. & 64$\pm$3 & 16. & 90$\pm$1 & 17. & {\color{blue}\textbf{81$\pm$1}} & 18. & 87$\pm$1 & 19. & 89$\pm$0
& \multicolumn{4}{l}{\ \textbf{Average: 82$\pm$1}} \\
\midrule
%\multirow{3}{0.18\linewidth}{Cross-entropy}
%& 1. & {\color{blue}\textbf{89$\pm$1}} & 2. & 51$\pm$5 & 3. & 75$\pm$1 & 4. & 84$\pm$1 & 5. & 75$\pm$2 & 6. & 67$\pm$2 & 7. & 75$\pm$3 \\
%& 8. & 58$\pm$2 & 9. & 64$\pm$1 & 10. & 84$\pm$1 & 11. & 88$\pm$1 & 12. & {\color{blue}\textbf{87$\pm$1}} & 13. & {\color{blue}\textbf{90$\pm$1}} & 14. & 75$\pm$1 \\
%& 15. & 56$\pm$2 & 16. & 89$\pm$1 & 17. & 74$\pm$1 & 18. & 86$\pm$1 & 19. & 86$\pm$1
%& \multicolumn{4}{l}{\ \textbf{Average: 76$\pm$0}} \\
%\midrule
%
\multicolumn{15}{c}{Gabor-based (1.53 million parameters)} \\
\midrule
\multirow{3}{0.18\linewidth}{$L_{PCC}$}
& 1. & 86$\pm$0 & 2. & 78$\pm$2 & 3. & 84$\pm$1 & 4. & 88$\pm$1 & 5. & 81$\pm$1 & 6. & 71$\pm$1 & 7. & 78$\pm$1 \\
& 8. & 68$\pm$2 & 9. & 72$\pm$2 & 10. & 84$\pm$1 & 11. & 87$\pm$1 & 12. & 85$\pm$1 & 13. & 88$\pm$0 & 14. & 78$\pm$1 \\
& 15. & 62$\pm$3 & 16. & 89$\pm$0 & 17. & 78$\pm$1 & 18. & 85$\pm$2 & 19. & 87$\pm$0
& \multicolumn{4}{l}{\ \textbf{Average: 80$\pm$0}} \\
\midrule
\multirow{3}{0.18\linewidth}{$L_{Dice}$}
& 1. & 83$\pm$1 & 2. & 78$\pm$1 & 3. & 84$\pm$0 & 4. & 88$\pm$1 & 5. & 80$\pm$1 & 6. & 72$\pm$1 & 7. & 77$\pm$2 \\
& 8. & 67$\pm$1 & 9. & 71$\pm$1 & 10. & 84$\pm$1 & 11. & 86$\pm$0 & 12. & 85$\pm$1 & 13. & 87$\pm$0 & 14. & 78$\pm$1 \\
& 15. & 62$\pm$2 & 16. & 89$\pm$0 & 17. & 78$\pm$1 & 18. & 85$\pm$1 & 19. & 86$\pm$0
& \multicolumn{4}{l}{\ \textbf{Average: 80$\pm$0}} \\
\midrule
%\multirow{3}{0.18\linewidth}{Cross-entropy}
%& 1. & 87$\pm$1 & 2. & 47$\pm$6 & 3. & 70$\pm$3 & 4. & 80$\pm$2 & 5. & 73$\pm$1 & 6. & 62$\pm$2 & 7. & 70$\pm$1 \\
%& 8. & 49$\pm$1 & 9. & 55$\pm$3 & 10. & 80$\pm$2 & 11. & 85$\pm$1 & 12. & 85$\pm$1 & 13. & 89$\pm$1 & 14. & 70$\pm$3 \\
%& 15. & 50$\pm$3 & 16. & 87$\pm$1 & 17. & 66$\pm$5 & 18. & 79$\pm$5 & 19. & 83$\pm$1
%& \multicolumn{4}{l}{\ \textbf{Average: 72$\pm$1}} \\
%\midrule
%
\multicolumn{15}{c}{Mixed-kernel (1.60 million parameters)} \\
\midrule
\multirow{3}{0.18\linewidth}{$L_{PCC}$}
& 1. & 87$\pm$1 & 2. & {\color{blue}\textbf{80$\pm$2}} & 3. & {\color{blue}\textbf{85$\pm$1}} & 4. & {\color{blue}\textbf{91$\pm$1}} & 5. & 83$\pm$0 & 6. & 75$\pm$1 & 7. & {\color{blue}\textbf{81$\pm$1}} \\
& 8. & 70$\pm$1 & 9. & {\color{blue}\textbf{77$\pm$1}} & 10. & 85$\pm$2 & 11. & {\color{blue}\textbf{90$\pm$0}} & 12. & {\color{blue}\textbf{87$\pm$1}} & 13. & 89$\pm$0 & 14. & {\color{blue}\textbf{82$\pm$1}} \\
& 15. & {\color{blue}\textbf{65$\pm$2}} & 16. & 90$\pm$1 & 17. & {\color{blue}\textbf{81$\pm$0}} & 18. & 87$\pm$1 & 19. & 89$\pm$0
& \multicolumn{4}{l}{\ {\color{blue}\textbf{Average: 83$\pm$0}}} \\
\midrule
\multirow{3}{0.18\linewidth}{$L_{Dice}$}
& 1. & 86$\pm$1 & 2. & 79$\pm$2 & 3. & {\color{blue}\textbf{85$\pm$1}} & 4. & {\color{blue}\textbf{91$\pm$0}} & 5. & 83$\pm$1 & 6. & {\color{blue}\textbf{76$\pm$1}} & 7. & 80$\pm$1 \\
& 8. & 70$\pm$1 & 9. & 75$\pm$1 & 10. & 84$\pm$2 & 11. & 89$\pm$0 & 12. & {\color{blue}\textbf{87$\pm$1}} & 13. & 88$\pm$0 & 14. & 81$\pm$0 \\
& 15. & 64$\pm$2 & 16. & 90$\pm$0 & 17. & 80$\pm$2 & 18. & 86$\pm$2 & 19. & 88$\pm$1
& \multicolumn{4}{l}{\ \textbf{Average: 82$\pm$0}} \\
%\midrule
%\multirow{3}{0.18\linewidth}{Cross-entropy}
%& 1. & 88$\pm$0 & 2. & 46$\pm$3 & 3. & 68$\pm$3 & 4. & 84$\pm$1 & 5. & 74$\pm$2 & 6. & 63$\pm$3 & 7. & 73$\pm$3 \\
%& 8. & 50$\pm$4 & 9. & 62$\pm$3 & 10. & 80$\pm$2 & 11. & 86$\pm$1 & 12. & 86$\pm$1 & 13. & {\color{blue}\textbf{90$\pm$1}} & 14. & 73$\pm$1 \\
%& 15. & 52$\pm$3 & 16. & 87$\pm$1 & 17. & 71$\pm$5 & 18. & 83$\pm$3 & 19. & 85$\pm$1
%& \multicolumn{4}{l}{\ \textbf{Average: 74$\pm$1}} \\
\bottomrule
\end{tabularx}
\end{table*}
% --------------------------------------------------

Fig. \ref{fig:learning_rate} shows the comparisons among the loss functions with respect to learning rates. Regardless of the kernel combinations, $L_{PCC}$ was the most robust and accurate one among the loss functions, while the categorial cross-entropy was also robust but less accurate. $L_{Dice}$ performed better than the categorical cross-entropy at learning rate $\leq$ 10$^{-3}$, but its performance dropped abruptly at larger learning rates. All loss functions had their performance decreased when the learning rates $<$ 10$^{-3}$, and the decrease of $L_{Dice}$ was more obvious than $L_{PCC}$. Comparing among different kernel combinations, the conventional models performed best in general, and the mixed-kernel models outperformed the Gabor-based models. Nevertheless, if we only concentrate on $L_{PCC}$, the conventional and mixed-kernel models had similar performance. They also had similar performance with $L_{Dice}$ at learning rates 10$^{-3.5}$ and 10$^{-3}$. Moreover, the conventional models were less tolerant to $L_{Dice}$ at larger learning rates.

As all loss functions performed well at learning rate 10$^{-3}$, the detailed comparisons at this rate are shown in Table \ref{table:brain}. Those of categorical cross-entropy are not shown because of their relatively low accuracy. The numbers of parameters of the conventional, Gabor-based, and mixed-kernel models were 4.99, 1.53, and 1.60 millions, respectively, thus the sizes of the conventional models were more than three times of the other models. If the kernel size of the conventional kernels changes from 3$\times$3$\times$3 to 5$\times$5$\times$5, i.e., the kernel size used by the V-Net \cite{Conference:Milletari:3DV2016}, the numbers of parameters of the conventional and mixed-kernel models become 22.84 and 1.95 millions, respectively, more than a ten-fold difference. Table \ref{table:brain} also shows that the conventional and mixed-kernel models performed similarly well with less than 1\% difference in Dice coefficients. The differences between using $L_{PCC}$ and $L_{Dice}$ were also less than 1\%. Furthermore, the overall framework was very robust to network initializations as the standard deviations from five dateset splits were less than 1\% on average. Note that although the Gabor-based models had the worst performance, they still had an average Dice coefficient of 80\% with the least numbers of parameters.

\begin{figure}[t]
    \fontsize{5}{6}\selectfont
    \centering
    \begin{minipage}[t]{0.13\linewidth}
      \centering
      \includegraphics[width=1\linewidth]{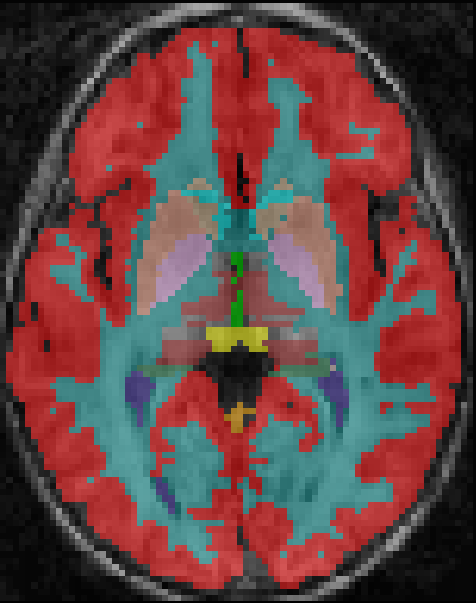} \\
      \includegraphics[width=1\linewidth]{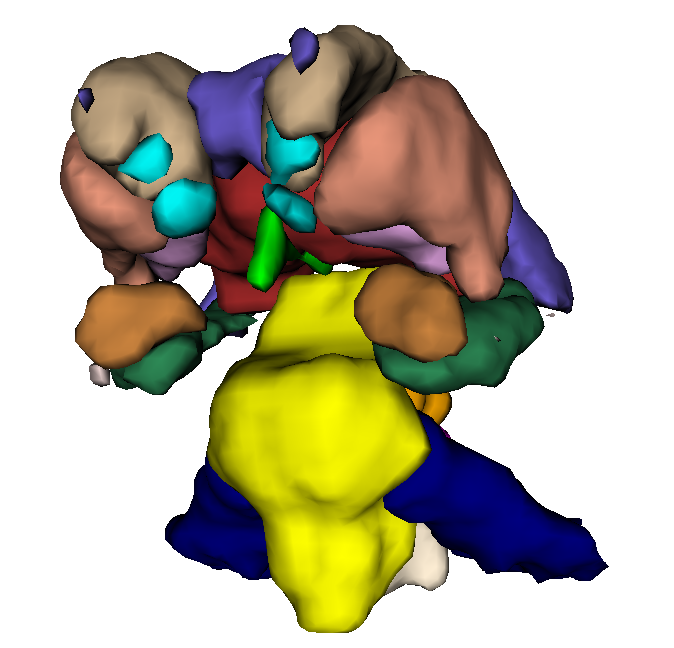} \\
      \centering{Ground truth}
    \end{minipage}
    \vrule\
    \begin{minipage}[t]{0.13\linewidth}
      \centering
      \includegraphics[width=1\linewidth]{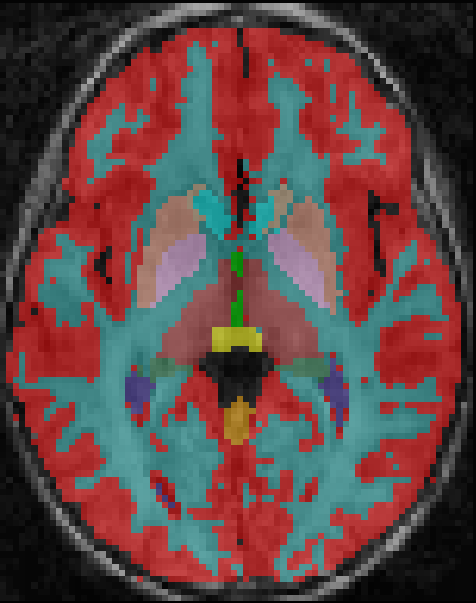} \\
      \includegraphics[width=1\linewidth]{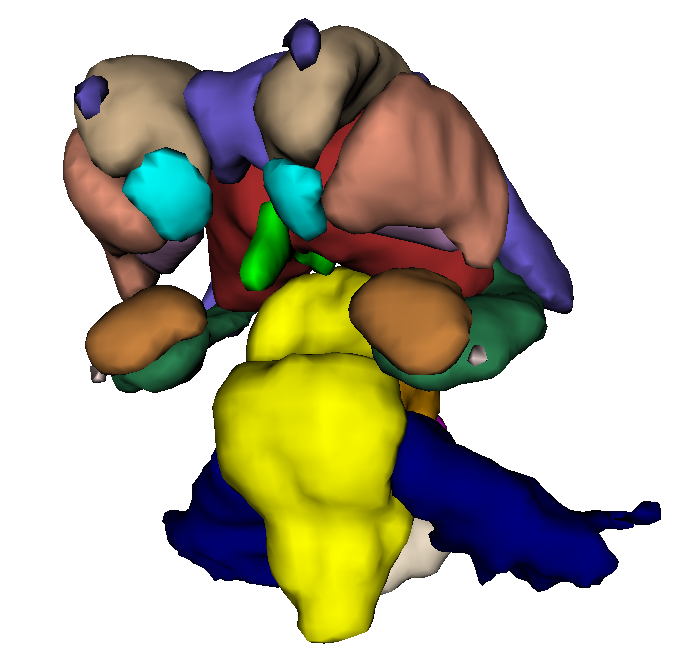} \\
      \centering{Conv: $L_{PCC}$ \\ Dice = 82\%}
    \end{minipage}
    \begin{minipage}[t]{0.13\linewidth}
      \centering
      \includegraphics[width=1\linewidth]{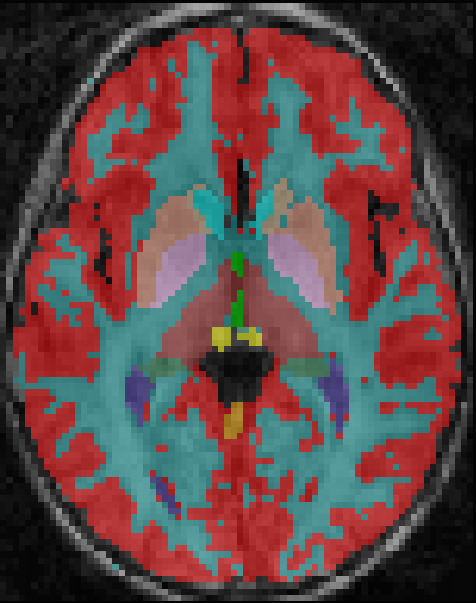} \\
      \includegraphics[width=1\linewidth]{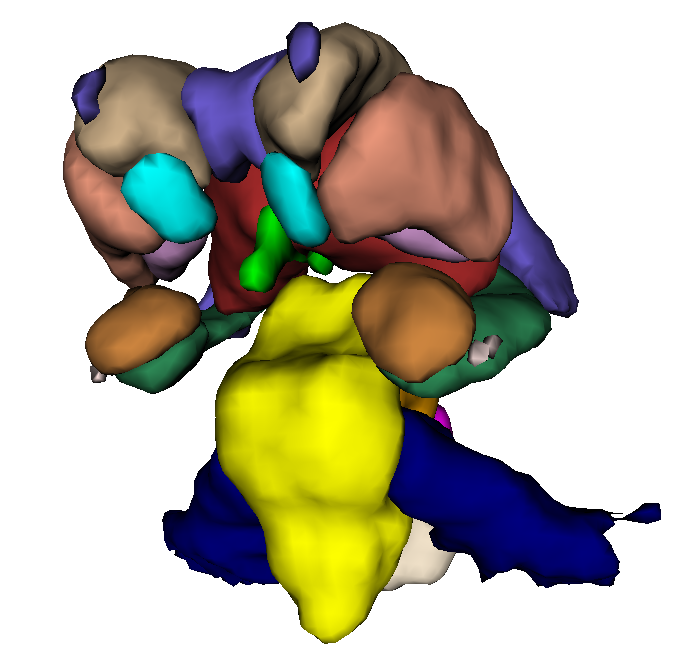} \\
      \centering{Conv: $L_{Dice}$ \\ Dice = 82\%}
    \end{minipage}
    \vrule\
    \begin{minipage}[t]{0.13\linewidth}
      \centering
      \includegraphics[width=1\linewidth]{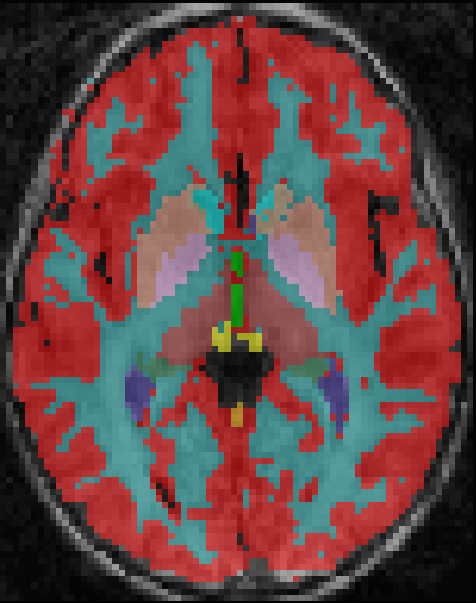} \\
      \includegraphics[width=1\linewidth]{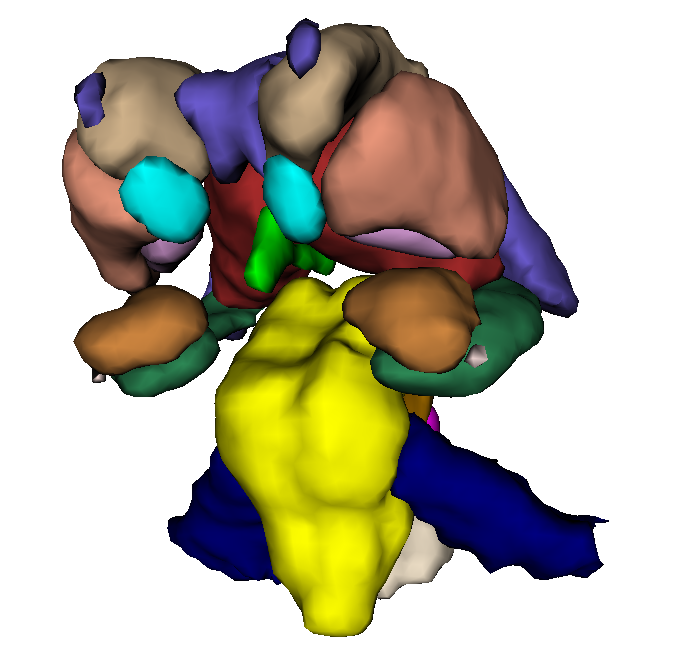} \\
      \centering{Gabor: $L_{PCC}$ \\ Dice = 80\%}
    \end{minipage}
    \begin{minipage}[t]{0.13\linewidth}
      \centering
      \includegraphics[width=1\linewidth]{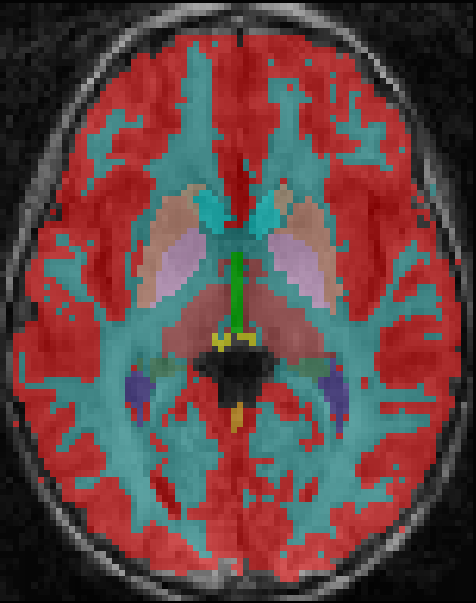} \\
      \includegraphics[width=1\linewidth]{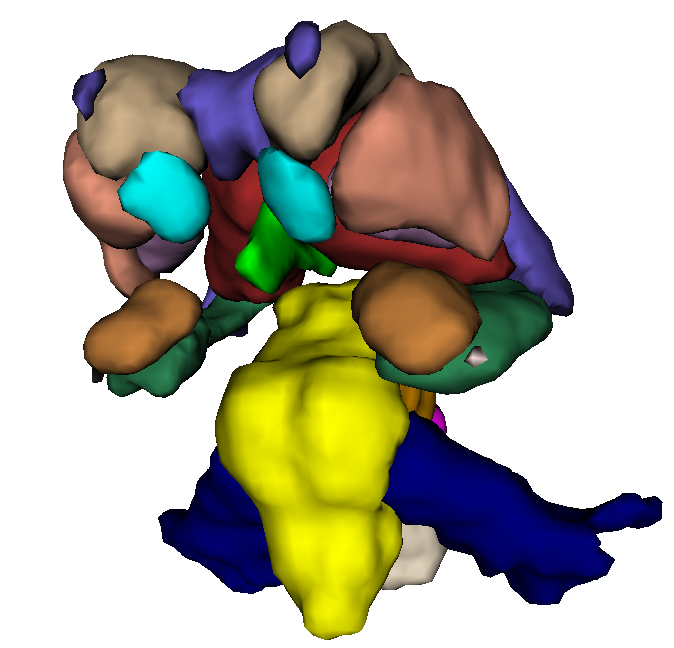} \\
      \centering{Gabor: $L_{Dice}$ \\ Dice = 79\%}
    \end{minipage}
    \vrule\
    \begin{minipage}[t]{0.13\linewidth}
      \centering
      \includegraphics[width=1\linewidth]{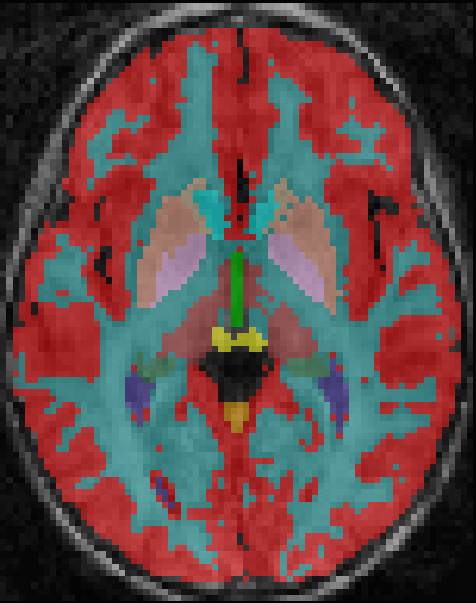} \\
      \includegraphics[width=1\linewidth]{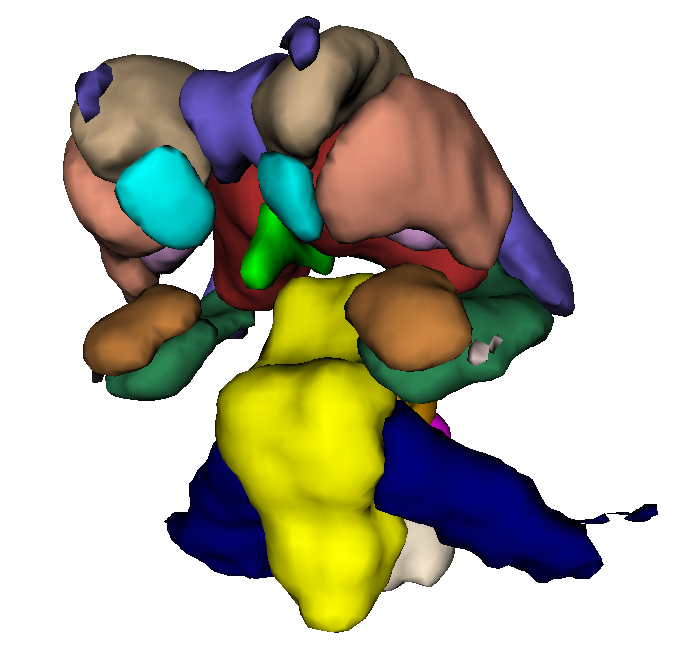} \\
      \centering{Mixed: $L_{PCC}$ \\ Dice = 82\%}
    \end{minipage}
    \begin{minipage}[t]{0.13\linewidth}
      \centering
      \includegraphics[width=1\linewidth]{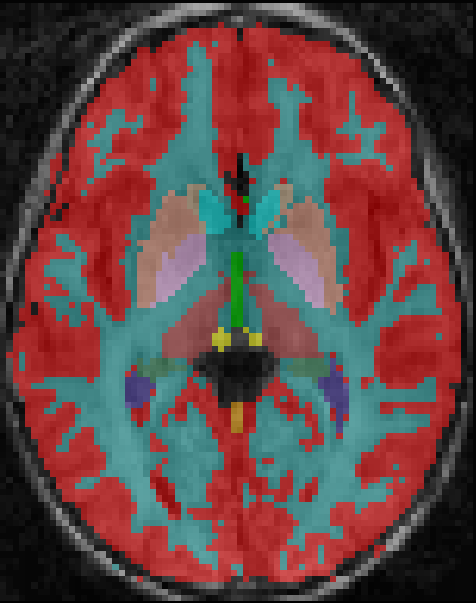} \\
      \includegraphics[width=1\linewidth]{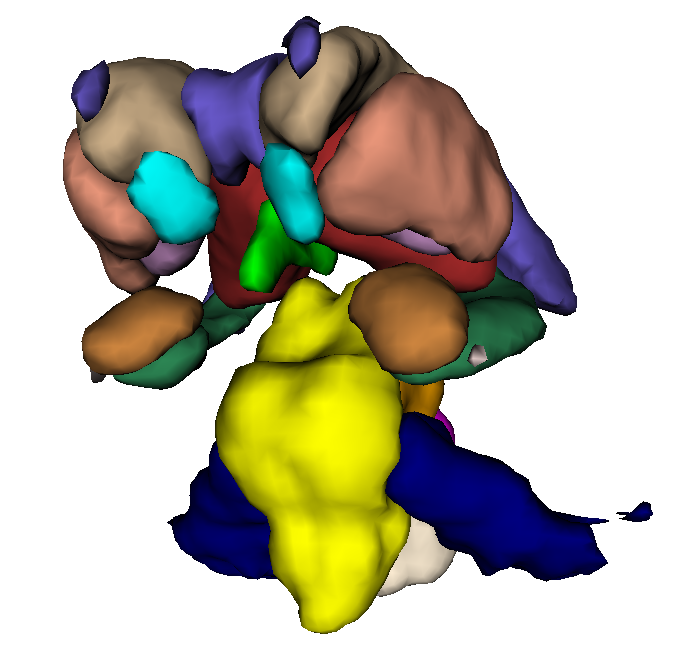} \\
      \centering{Mixed: $L_{Dice}$ \\ Dice = 82\%}
    \end{minipage}
    \caption{Visualization of an example. Top: axial view. Bottom: 3D view with the cerebral grey, cerebral white, and cerebellar grey matters hidden for better illustration.}
    \label{fig:visualization}
\end{figure}

% Visualization of an example
Fig. \ref{fig:visualization} shows the visualization of an example. Although the Dice coefficients of the Gabor-based models were 2\% to 3\% lower than the other models, their segmentations were very similar to the ground truth.

From the experimental results, we learn that $L_{PCC}$ was more robust than $L_{Dice}$ and more accurate than the categorical cross-entropy. The accuracies of different kernel combinations were very similar especially between the conventional and mixed-kernel models, but the mixed-kernel models used much fewer numbers of parameters. Such differences in size can be more obvious if larger kernel sizes are used. Although the Gabor-based models had the worst performance among the tested models, they still provided an average Dice coefficient of 80\%. This is a good demonstration that parametric kernels can be learned through backpropagation in deep learning with decent performance.

% ------------------ figure -------------------------
% ---------------------------------------

\section{Conclusion}

In this paper, we propose a fully trainable Gabor-based kernel and a loss function based on the Pearson's correlation coefficient. Experimental results show that $L_{PCC}$ is robust to learning rate and can achieve high segmentation accuracy, and proper combinations of conventional and Gabor-based kernels can result in accurate models that are multiple times smaller than the conventional models.

%\newpage
\bibliographystyle{splncs03}
\bibliography{Ref}

\end{document}